\documentstyle[12pt,fleqn,epsfig]{article}
\begin{document}

\begin{center}
{\bf INSTITUT~F\"{U}R~KERNPHYSIK,~UNIVERSIT\"{A}T~FRANKFURT}\\
D - 60486 Frankfurt, August--Euler--Strasse 6, Germany
\end{center}

\hfill IKF--HENPG/1--98


\vspace{0.5cm}
\begin{center}
   {\Large \bf On Transverse Momentum Event--by--Event}  
\end{center}
\begin{center}
   {\Large \bf Fluctuations in String Hadronic Models }
\end{center}

\vspace{1cm}
\begin{center}

Feng Liu\footnote{E--mail: fliu@iopp.ccnu.edu.cn}\\
Institut f\"ur Kernphysik, Universit\"at Frankfurt,
Germany and\\
Institute of Particle Physics, Huazhong Normal University,
Wuhan, China\\[0.8cm]

An Tai\footnote{E--mail: taian@hptc1.ihep.ac.cn}\\
Institute of High Energy Physics, Beijing, China
\\[0.8cm]

Marek Ga\'zdzicki\footnote{E--mail: marek@ikf.physik.uni--frankfurt.de}
~~~~and~~~~ 
Reinhard Stock\footnote{E--mail: stock@ikf.physik.uni--frankfurt.de}\\
Institut f\"ur Kernphysik, Universit\"at Frankfurt,
Germany\\[0.8cm]

\end{center}

\vspace{0.5cm}

\begin{abstract}\noindent

Transverse momentum event--by--event fluctuations are studied within
the string--hadronic model of high energy nuclear collisions, LUCIAE.
Data on 
non--statistical $p_T$ fluctuations in p+p interactions are reproduced.
Fluctuations of similar magnitude are predicted for nucleus--nucleus
collisions, in contradiction to the preliminary NA49 results. 
The introduction of a string clustering mechanism (Firecracker Model)
leads to a further, significant increase of $p_T$ fluctuations 
for nucleus--nucleus
collisions.
Secondary hadronic interactions, as implemented in LUCIAE, 
cause only a small
reduction of $p_T$ fluctuations.

\end{abstract}

\newpage

\section{Introduction}

Experimental results on event--by--event fluctuations in nuclear
collisions at relativistic energy
may serve as a crucial test of various theoretical approaches to the
collision process. Until now the existing models were developed with the
aim to reproduce 
the data on single particle yields and two particle correlations.
To which extend can the same models  {\it predict} 
event--by--event fluctuations?  

A method to study  fluctuations of global kinematical observables in high
energy nuclear collisions (like fluctuations of total transverse momentum)
was proposed in \cite{Ga:92}. 
It was applied to analyze experimental data
\cite{Ro:97, Ch:98} and test theoretical approaches 
\cite{Ga:97, Bl:98, Mr:98}.
The method is based on the fact that in elementary interactions (e.g. p+p 
interactions) particles are produced in a correlated way which leads to the
observation of large (non--statistical) event--by--event fluctuations.
These 'elementary' fluctuations provide a scale relative to which
the fluctuations in nuclear collisions can be studied in a model 
independent way.
The method also provides  a statistical tool to account for `trivial'
geometrical fluctuations.

Recently it was shown that initial state scattering models lead to
increase of $p_T$ fluctuations \cite{Ga:97}.
This is in contradiction with the experimental results which indicate
that the properly normalized $p_T$ fluctuations in central Pb+Pb collisions
at 158 A$\cdot$GeV are significantly smaller than the corresponding 
fluctuations in p+p interactions \cite{Ro:97,Ad:98}.
To which level can the experimental result be understood as an effect of
equilibration due to  hadronic cascading?
What is the role of  possible collective effects at the early stage of the collision,
like clustering of strings?
The aim of this paper is to investigate  these questions using the string--hadronic
model LUCIAE \cite{luciae}. 

The paper is organized as follows.
The analysis method used further to study event--by--fluctuations is recalled
in Section 2.
In the Section 3 we briefly sketch the basic physics ingredients of the LUCIAE.
Model calculations concerning $p_T$ fluctuations are presented 
and discussed in Section 4.
Summary and discussion close the paper.

\section{A Measure of Event-by-Event Fluctuations}

Event--by--event fluctuations in nuclear collisions are usually dominated by
the trivial variation in impact parameter from event to event and 
by the purely
statistical (here we mean statistics for classical particles) 
variation of the measured quantities. 
An analysis method that
allows  to remove  these trivial contributions and to  determine the
remaining part of event--by--event fluctuations 
of transverse momentum has been proposed in \cite{Ga:92}.
 Following this reference we define for every particle $i$ in an event:
\begin{eqnarray*}
z_i = p_{T_i} - \overline{p_T},
\end{eqnarray*}
where $\overline{p_T}$ is the mean transverse momentum of accepted particles
averaged over all events (the inclusive mean).
Using $z_i$ we calculate for every
event
\begin{eqnarray*}
Z = \sum_{i=1}^N z_i,
\end{eqnarray*}
where $N$ is the number of accepted particles in the event. 
With this definition
one obtains the following measure of event--by--event fluctuations:
\begin{eqnarray}
\Phi_{p_T}  = \sqrt{\frac{\langle Z^2 \rangle}{\langle N \rangle}} -
\sqrt{\; \overline{z^2}},
\label{Phi}
\end{eqnarray}
where $\langle N \rangle$ and $\langle Z^2 \rangle$ are averages over all
events
and the second term in the r.h.s. is the square root of the second moment
of the inclusive transverse momentum distribution.\\
The physical motivation for studying $\Phi_{p_T}$ was given in \cite{Ga:92}:
experimental data on N+N interactions show
that particles in these collisions are not produced independently \cite{Go:84}.
One observes large scale correlations that lead to, e.g.,\ a correlation
between the event multiplicity and the average $p_T$ of the particles.
The correlated particle emission in elementary processes can
be used to probe the dynamics of nuclear collisions
by
measuring to which degree this correlation 
is 
changed when going to p+A and/or A+A collisions.\\
For this purpose, $\Phi_{p_T}$ as a measure of
fluctuations has two important properties. For a large
system (i.e.\ an A+A collision) that is a superposition of
many independent elementary systems (i.e.\ N+N interactions),
$\Phi_{p_T}$ has a constant value that is identical to that
of the elementary system.
In other words if the central  Pb+Pb collisions were an 
independent superposition
of N+N interactions, the value of $\Phi_{p_T}$ would
remain constant, independent of the number of
superimposed elementary interactions
in a single event and its distribution in the studied
sample of the events.
If on the other hand the large system consists of particles
that have been emitted independently, $\Phi_{p_T}$ assumes
a value of zero.
Thus $\Phi_{p_T}$  provides us with a scale
characterising the fluctuations in nuclear collisions
relative to elementary interactions at the same
energy.\\
One should expect that $\Phi_{p_T}$ is sensitive to 
both event--by--event fluctuations in the creation of the
early state of the collision as well as in its subsequent evolution until
freeze--out.\\

\section{The string--hadronic model -- LUCIAE}

The LUCIAE model is developed based on  FRITIOF  \cite{fritiof}. 
In the FRITIOF model a hadron is assumed to behave 
like a massless relativistic string corresponding to a confined color 
force field of a vortex line character embedded in a type II color 
superconducting vacuum.  
In FRITIOF, 
during the collision two hadrons are excited due to longitudinal momentum 
transfers and/or  Rutherford Parton Scattering (RPS). The highly excited 
states will emit bremsstrahlung gluons according to the soft radiation model. 
They are afterwards treated as excitations, i.e. the Lund Strings, and  
allowed to decay into final state hadrons according to the Lund
fragmentation scheme \cite{lund}.

The LUCIAE  
includes all elements of the FRITIOF model and, additionaly, two
components: a `Firecracker' model and 
a model of hadron rescattering.
In the Firecracker model it is assumed that  groups of neighbouring
strings  may form interacting quantum states (clusters) so that both the
emission of gluonic bremsstrahlung as well as the fragmentation
properties can be affected by the large common energy density of
the interacting strings. 
The maximum transverse momentum 
of the emitted gluons is found to fulfill a condition \cite{fire}
$k_{\perp max} \leq \sqrt{\mu M_{tot}}$, where $M_{tot}$ is the total excitation
energy of a cluster and $\mu M_{tot}$  effectively corresponds 
to an energy density over a region of  transverse size $1/\mu$.
Consequently, when it comes to heavy ion collision predictions the Firecracker
model  will
correspond to an essential enhancement of (mini)jets in the center of
phase space, which contributes to high $p_T$ enhancement \cite{fire}.

In the rescattering model, the produced
particles (which consist of
hadrons after strong decays  
and the participant nucleons) are distributed 
randomly in the  geometrical overlapping region
between the projectile and the target nuclei.  
The target (projectile) spectator nucleons are distributed randomly 
outside the
overlapping region and inside the target (projectile) sphere. 
A formation time is given to each particle
and a particle starts to scatter with others after it is
``formed''. Two particles will collide if their minimum distance 
$d_{min} \leq \sqrt{\sigma_{tot}/\pi},$ where $\sigma_{tot}$
 is the total cross section in fm$^2$ and the 
minimum distance is calculated in the cms frame of the two
colliding particles (for details see \cite{luciae}).  
\newpage

\section{The $p_T$ Fluctuations in LUCIAE}

The analysis of the event--by--event fluctuations in LUCIAE
is done in several steps.
In the first part of this section we study event--by--event
fluctuations in p+p interactions.           
In the following subsections the $p_T$
fluctuations in A+A collisions are analyzed.
We study effects of independent string superposition,
string clustering and finally the role of the hadronic rescattering.

In the analysis presented  in this paper  the $\phi_{p_T}$ variable
was calculated using all charged particles stable with respect to
strong interaction, charged spectator fragments  were excluded.
The number of generated events for all analyzed event samples
was larger than $10^4$. 
The statistical error of the $\phi_{p_T}$ variable
was calculated according to  formula given in the
Appendix.

\subsection{Fluctuations in p+p Interactions}

It is well established experimenatlly 
\cite{Go:84} that  particle production
in p+p interactions at high energy is correlated.
It was pointed out in Ref. \cite{Ga:92}
that this correlation could lead to        
a large non--statistical event--by--event fluctuations which
can be quantified by the $\Phi$ measurement.

It is essential that the string--hadronic models which are used 
for the analysis of the event--by--event fluctuations in
A+A collisions are first checked as to whether they reproduce
the corresponding fluctuations measured in p+p interactions.
This is because in these models the latter process is used
as an input for the calculation of the properties of the
A+A collisions.

It was shown \cite{Ga:92} that the  $\Phi_{P_T}$ variable is sensitive
to the correlation between the form of the $p_T$ distribution
and the event multiplicity.
Therefore in Fig. 1 we compare the experimental data on
$\langle p_T \rangle$ vs $n$ dependence for p+p interactions
at 200 GeV \cite{Ka:77} with the LUCIAE results.
The comparison is done separately for $\pi^+$ and $\pi^-$
mesons. 
Two versions of LUCIAE are used.
In Figs. 1a and 1c the results for the full version of the
model, which includes hard processes (RPS), are shown.
The results obtained by switching off hard processes in the model
are presented in Figs. 1b and 1d.
We note here that the string clustering model as well as hadronic
rescattering play no role for p+p interactions.

The comparison indicates that LUCIAE underpredicts both
the absolute magnitude and the strength of 
the $\langle p_T \rangle$ vs $n$ dependence. 
The hard scattering  processes, which occur in about 5\% of
p+p interactions at 200 GeV, play a minor role when
the $\langle p_T \rangle$ vs $n$ dependence is considered.
 
The fact that   
the $\langle p_T \rangle$ vs $n$ dependence 
is underpredicted by  LUCIAE may suggest that
the event--by--event $p_T$ fluctuations  are smaller
in the model than in the data.
This is however not the case.
The $\Phi_{p_T}$ value calculated within LUCIAE without
hard processes 
is about 15 MeV/c for p+p interaction
at 158 GeV.
This value of $\Phi_{p_T}$ agrees with the corresponding 
preliminary experimental value
obtained by the NA49 Collaboration \cite{Ad:98}
when the acceptance effects are taken into account.
The introduction of  hard scattering increases the $\Phi_{P_T}$ value
to 30 MeV/c leading to an overestimation of the
experimental number.

The above results illustrate two important features of $\Phi_{p_T}$.
Its value is determined not only by the $\langle p_T \rangle$ vs $n$
dependence but also by other correlations present in the particle
production process (e.g. jet production).
It is sensitive to different properties of the production process
than the inclusive or semi--inclusive observables and  therefore
yields additional information otherwise not available.

The  strong increase of the $\Phi_{p_T}$ value when the hard processes
are included may be understood as due to the fact that the final
state particles originating from the hard process (jets) are
strongly correlated in momentum space.

In order to trace the origin of the particle correlation
in the p+p interactions for soft processes in LUCIAE 
we exchanged the string fragmentation scheme by an independent 
fragmentation scheme in which energy--momentum and quantum charges
are not conserved.
This results in a reduction of 
the $\Phi_{p_T}$ value by a factor of about 5, to almost zero.
Thus we conclude that the conservation laws are responsible for  large
event--by--event fluctuations in soft p+p interactions at SPS energies in
LUCIAE.

For the further study of $p_T$ fluctuations in A+A collisions
we selected the LUCIAE version without hard processes as only this 
version reproduces correctly the magnitude of the $p_T$ 
fluctuations measured for p+p interactions \cite{Ad:98}.

\subsection{Superposition of Strings}

The fundamental assumption of the string--hadronic models
of A+A collisions is that the basic physics of these
collisions can be pictured as an (almost) independent
superposition of nucleon--nucleon (N+N) interactions.

Thus in order to study properties of this minimal model
with respect to $p_T$ fluctuations we analyze the fluctuations
calculated within the LUCIAE without hard processes, string
clustering and hadronic rescattering.
For this version of the model we calculate the $\Phi_{p_T}$
for central ($b = 0$ fm) A+A collisions and
plot it as a function of $A$ in Fig. 2.

One observes a  independence of  $\Phi_{p_T}$
of $A$.   
This behaviour is  expected, by definition, 
for a model in which A+A is assumed to be an independent
superposition of N+N interactions.
In the studied version of LUCIAE there are two effects which
can cause deviations from the independent superposition picture.
The nucleons in the nucleus have Fermi motion and
the string excitation increases with the size of the
colliding nuclei.
Our numerical results show that both effects play a minor role
when the $\Phi_{p_T}$ is studied.
Therefore, in this respect, the minimal version of the
string--hadronic
model of A+A collisions can be considered as an independent superposition 
of N+N interactions.

\subsection{String Clustering}

In the high density stage of the A+A collisions 
the picture of independent string superposition 
is obviously unrealistic.
A modification of this picture
is modeled, in the framework of string--hadronic
models, by introduction of string clustering
(Firecracker model) \cite{luciae} or string fusion \cite{fusion}.

In the Firecracker model clustering of strings allows
for a collective  gluon emission and therefore
is expected to increase the $p_T$ fluctuations.
The probability that the string will form a cluster 
increases with the size of the colliding nuclei
and consequently the $p_T$ fluctuations
due to string clustering
should increase with the size of the colliding nuclei.
In fact this is observed in Fig. 2, where the results 
of the calculations with string clustering effect are shown.
For central Pb+Pb collisions the value of $\Phi_{p_T}$
increases by a factor of about 3 when the string clustering
is added to the minimal version of the LUCIAE model.

\subsection{Rescattering}

The
early stage of A+A collisions is followed
by the stage in which hadrons and hadronic resonances are effective
degrees of freedom.
Considerable rescattering between them is expected to take place before the
final decoupling of produced particles.
For the interpretation of the measured event--by--event fluctuations
it is crucial to understand the role played by the hadronic
rescattering process.

The  $\Phi_{p_T}$ calculated for central S+S and Pb+Pb collisions
within the minimal version of the 
LUCIAE supplemented by the hadronic rescattering model
is shown in Fig. 2.
One observes that the $p_T$ fluctuations are slightly reduced
by the hadronic secondary interactions.
This trend can be, in fact, 
understood as the rescattering should lead to the
equilibration of the final state and therefore it should reduce
the fluctuations established at the early stage.
From that point of view it may even be surprising that the
role of the rescattering is relatively small,
as each produced hadron rescatters in average 6 timesi until
decoupling in central Pb+Pb collision modeled by the present
version of the LUCIAE.

\newpage

\section{Summary and Discussion}

The analysis of event--by--event fluctuations in string--hadronic
model LUCIAE presented in this paper was triggered by the experimental
results of the NA49 Collaboration \cite{Ro:97,Ad:98}. 
They indicate that $p_T$ fluctuations, as measured by 
$\Phi_{p_T}$ measure, are significantly reduced in central
Pb+Pb collisions at 158 A$\cdot$GeV in comparison to p+p interactions
at the same energy.

\noindent
Our main results can be summarized as follows.
\begin{enumerate}
\item
\noindent
The $\Phi_{p_T}$ value in p+p interactions modeled by
LUCIAE is greater than zero.
This is caused by correlations between particles introduced
by conservation laws and by hard scattering.
\noindent
\item
The $\Phi_{p_T}$ value is independent of the size of the colliding
nuclei  when A+A collisions are modeled by the minimal version
of the LUCIAE model (without hard scattering, 
string clustering and hadronic rescattering).
It remains constant at the value found for p+p interactions.
\item
\noindent
A string clustering effect, as introduced in LUCIAE, causes a strong
increase of  $\Phi_{p_T}$ with  increasing size of the 
colliding nuclei.
\item
Final state hadronic rescattering, as modeled in
LUCIAE, only weakly decreases the value
of $\Phi_{p_T}$.
\end{enumerate}

Our results should be confronted with the recently presented 
results on event--by--event fluctuations as measured by the $\Phi_{p_T}$.  

It was found in \cite{Ga:97} that  initial state scattering
models predict an increase of $\Phi_{p_T}$ with  the
size of the colliding nuclei.
This behaviour is similar to the behaviour obtained by us for
the string clustering effect.
The relative contribution of both effects (initial state scattering
and string clustering) increases with the size of the colliding nuclei.

The
influence of  hadronic rescattering on event--by--event fluctuations
was studied in Ref. \cite{Ch:98} using the VENUS model \cite{We:95}
and in Ref. \cite{Bl:98} using the UrQMD model 
\cite{UrQMD}.
The secondary interactions as modelled by VENUS were shown not to
influence fluctuations in pseudorapidity measured by $\Phi_{\eta}$
\cite{Ch:98}.
UrQMD finds a strong reduction of the value of $\Phi_{p_T}$ when
going from p+p interactions to central Pb+Pb collisions.
This result is in contradiction with our finding (see point 4 above).  
We are aware of the following differences between the current
analysis and the analysis done within UrQMD model \cite{Bl:98,Bl:pr}.
In the UrQMD analysis all particles at midrapidity
($-0.5<y^*<0.5$) are used whereas in our analysis only  charged
particles but without rapidity selection are included.
We checked however that our conclusion on the weak influcence of rescattering
remains unchanged when a `UrQMD acceptance' is used for LUCIAE events.
The number of central Pb+Pb events with rescattering was 10000
for LUCIAE and 1600 for UrQMD\footnote{In both models generation of
a single Pb+Pb event takes about 30' CPU time  on Pentium Pro 200.}.
The rescattering prescriptions are different in UrQMD and LUCIAE.
We checked however that the number of secondary hadronic collisions
in central Pb+Pb collision is similar (about 6 rescatterings
per final state hadron). 

The difference in the conclusions 
may be, also, due to the fact
that in UrQMD model Pauli blocking of baryons is taken into account.
This leads us to a comment on the results obtained recently by
Mr\'owczy\'nski \cite{Mr:98}.
He calculated the value of $\Phi_{p_T}$ for an equilibrium ideal gas in
the grand canonical approximation.
In the case of classical particles  $\Phi_{p_T}$ = 0.
The $\Phi_{p_T}$ value for fermions is large negative, whereas
for bosons large positive.
He estimates that due to the dominance of pions in the final state
the value of $\Phi_{p_T}$ in the hadronic gas should be large
and positive and therefore it is not easy to understand the NA49 results
even  assuming full equilibration of the matter.

We conclude from the above sketchy review 
that the basic discrepancy
between the results of various models remains to be
reconciled.
Vis a vis the preliminary data obtained by NA49 \cite{Ro:97,Ad:98}
which indicate a vanishing $\Phi_{p_T}$ measure both the results
of our microscopic  string--hadronic model LUCIAE, 
and expectation based on an equilibrium hadron quantum gas
miss the mark.
On the other hand the UrQMD model which is based 
on a similar physical picture predicts a vanishing 
$\Phi_{p_T}$ measure.
Further, ongoing work is devoted to pinpoint the 
origin(s) of this apparent discrepancy.
We also urgently await final affirmation of the
preliminary NA49 data.

\vspace{1cm}
\noindent
{\bf Acknowledgements}

We would like to thank H. Str\"obele and G. Roland 
for discussion and comments.
This work was partly supported by DFG (Germany) and
National Natural Science Fundation (China).

\newpage

\noindent
{\Large \bf Appendix}

\vspace{0.5cm}

{\bf A.} The fluctuation measure $\Phi_x$, where $x$ is any single
particle variable, defined by Eq.~1 can be expressed
in an explicit way by single event variables
$N$, $X$ and $X_2$, where N is the number of particles and  
$$
X = \sum_{i=1}^N x_i \;,\;\;\;\; 
X_2 = \sum_{i=1}^N x_i^2.
$$

It is easy to show that the definition (Eq.~1) is 
equivalent to
\begin{equation}
\Phi_x = \bigg( \frac {\langle X^2 \rangle} {\langle N \rangle} -
 \frac {2 \langle X \rangle \langle X N \rangle} {\langle N \rangle^2} +
 \frac {\langle N^2 \rangle \langle X \rangle^2} {\langle N \rangle^3} \bigg)^{1/2} -
 \bigg( \frac {\langle X_2 \rangle} {\langle N \rangle} -
 \frac {\langle X \rangle^2} {\langle N \rangle^2} \bigg)^{1/2}.
\end{equation} 
The above formula allows to calculate value of $\Phi_x$ during a single pass
of data processing, without initial evaluation of the
inclusive mean, $\overline{x}$.

\vspace{0.2cm}

It is common to study event--by--event fluctuations by the analysis
of the variance of the ratio $X/N$ (the mean $x$ for particles from a
given event).
From the definition of variance we get:
\begin{equation}
V(\frac {X} {N}) = \langle \frac {X^2} {N^2} \rangle -
                   \langle  \frac {X} {N} \rangle^2.
\end{equation}
We make a trivial observation that the $\Phi_x$ and $V(X/N)$
are different functions of different moments of basic single
event observables $X$ and $N$.
Thus analysis of $\Phi_x$ is not equivalent to the analysis
of $V(X/N)$.

\vspace{0.5cm}

{\bf B.} One can estimate a statistical 
error of the $\Phi_x$, $\sigma(\Phi_x)$,
when the single event variables $N$, $Z^2$ and $Z_2$ 
($Z_2 = \sum_{i=1}^N z_i^2$) are considered
as the original random variables.
The $\Phi_x$ can be then written as:
\begin{equation}
\Phi_x = \bigg(\frac {\sum_{j=1}^{N_{EV}} Z_j^2}  {\sum_{j=1}^{N_{EV}} N_j}\bigg)^{1/2} -
         \bigg(\frac {\sum_{j=1}^{N_{EV}} Z_{2,j}}  {\sum_{j=1}^{N_{EV}} N_j}\bigg)^{1/2},
\end{equation}
where $N_{EV}$ is the number of events.
The $\sigma(\Phi_x)$ can be therefore expressed as:
\begin{eqnarray*}
\frac {\sigma^2(\Phi_x)} {N_{EV}}  &=&
\bigg(\frac {\delta \Phi_x} {\delta Z^2}\bigg)^2 V(Z^2, Z^2) +
\bigg(\frac {\delta \Phi_x} {\delta Z_2}\bigg)^2 V(Z_2, Z_2) + 
\bigg(\frac {\delta \Phi_x} {\delta N}\bigg)^2 V(N, N) +  \\
 & & \frac {\delta \Phi_x} {\delta Z^2} \frac {\delta \Phi_x} {\delta N} V(Z^2, N) + 
\frac {\delta \Phi_x} {\delta Z_2} \frac {\delta \Phi_x} {\delta Z^2} V(Z_2, Z^2) + 
\frac {\delta \Phi_x} {\delta N} \frac {\delta \Phi_x} {\delta Z_2} V(N, Z_2), 
\end{eqnarray*}
where $\delta \Phi_x / \delta (N, Z^2, Z_2)$
is a derivate of $\Phi_x$ over a single event variable ($N, Z^2, Z_2$)
and $V(Y_1, Y_2) = \langle (Y_1 - \langle Y_1 \rangle)(Y_2 - \langle Y_2 \rangle) \rangle$.
Of course finally the expression for the statistical error of $\Phi_x$ can be
written in terms of the algebraic moments  as in the case of the
observable $\Phi_x$ itself (Eq.~2).

\newpage

\begin{figure}[p]
\epsfig{file=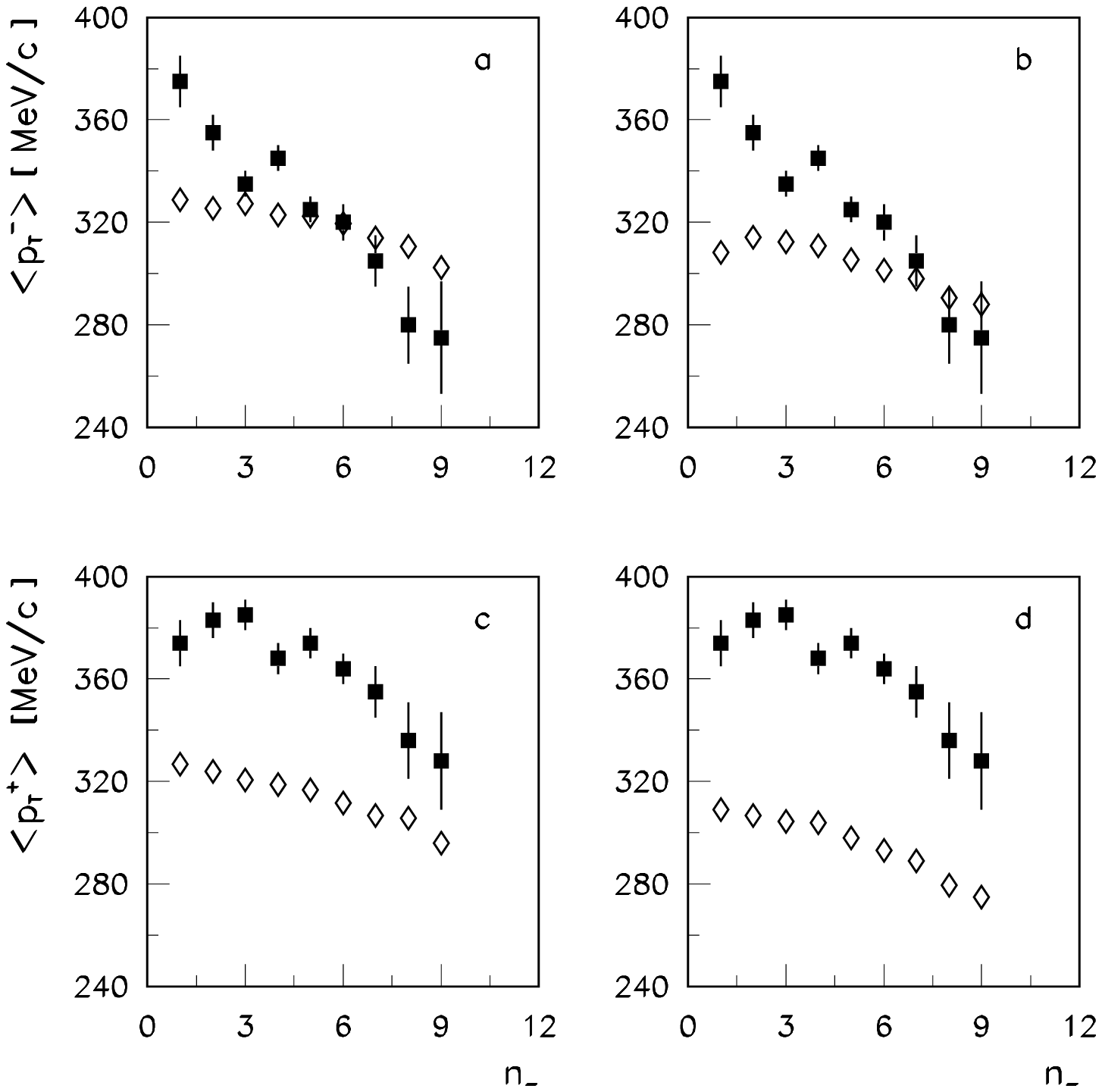,width=14cm}
\caption{
The dependence of the average transverse momentum of
$\pi^-$ (Figs. 1a and 1b) and $\pi^+$ 
(Figs. 1c and 1d) mesons on multiplicity of negatively
charged hadrons for p+p interactions at 200 GeV.
The experimental results \protect{\cite{Ka:77}}
are indicated by the filled squares,
whereas the results obtained within the LUCIAE
model by the open diamonds.
The hard processes are included in the case of Figs. 1a and
1c, and excluded in the case of Figs. 1b and 1d.
}
\label{fig1}
\end{figure}

\newpage

\begin{figure}[p]
\epsfig{file=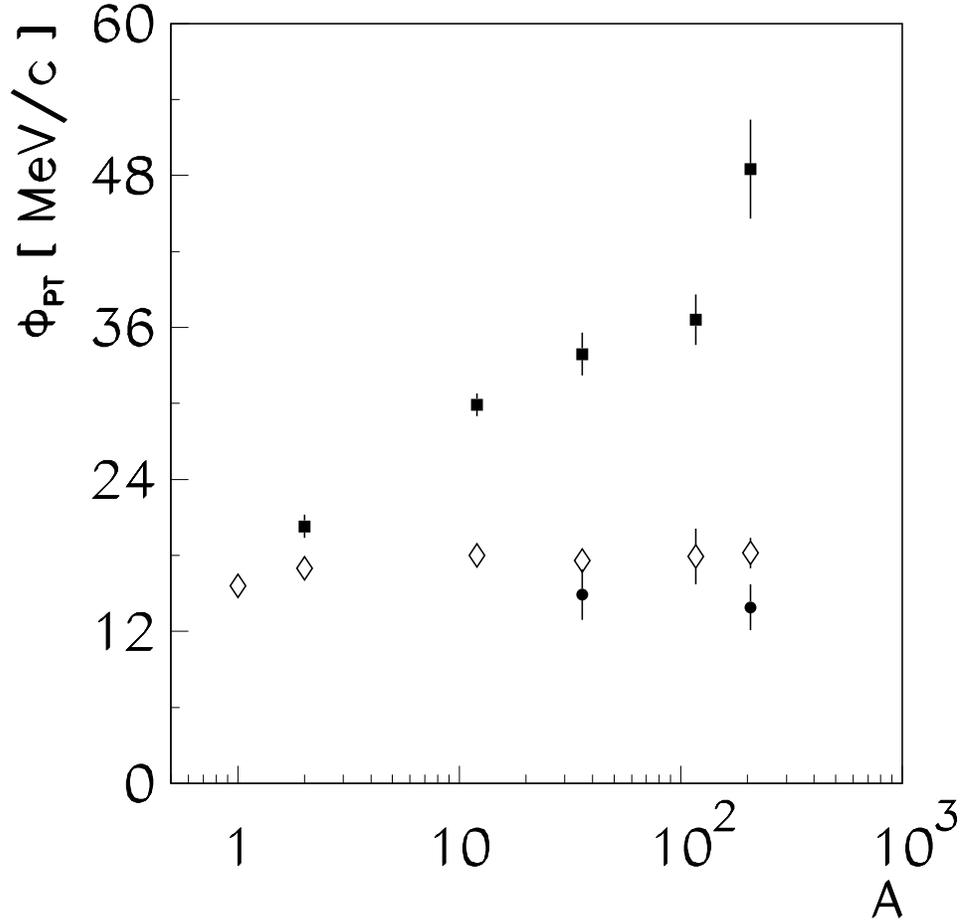,width=14cm}
\caption{
The dependence of the $\Phi_{p_T}$ on the
nuclear mass number of the colliding nuclei
calculated for A+A collisions at 158 A GeV at zero impact parameter
using LUCIAE model. 
Results obtained without hard scattering, string clustering
and  secondary hadronic
interactions (minimal version of the model) are indicated by open diamonds. 
Behaviour obtained by including string clustering
is shown by filled squares.
Filled circles indicate the value of $\Phi_{p_T}$ obtained
when the hadronic rescattering is added to the minimal version of
the model.
}
\label{fig2}
\end{figure}

\end{document}